\documentclass[a4paper,11pt]{article}

\usepackage{jcappub} 

\usepackage[T1]{fontenc} 

\title{\boldmath Ultra high energy cosmic rays from super-heavy dark matter in the context of large exposure observatories}

\author[a,1]{A. D. Supanitsky\note{Corresponding author.}}
\author[b]{and G. Medina-Tanco}

\affiliation[a]{Instituto de Tecnolog\'ias en Detecci\'on y Astropart\'iculas (CNEA, CONICET, UNSAM),\\
Centro At\'omico Constituyentes, San Martin, Buenos Aires, Argentina.}
\affiliation[b]{Instituto de Ciencias Nucleares, UNAM, \\ Circuito Exteriror S/N, Ciudad Universitaria,
M\'exico D. F. 04510, M\'exico.}

\emailAdd{daniel.supanitsky@iteda.cnea.gov.ar}
\emailAdd{gmtanco@nucleares.unam.mx}

\abstract{The origin of the ultra high energy cosmic rays (UHECRs, $E>10^{18}$ eV) is still uncertain. However, great
progress has been achieved due to the data taken by The Pierre Auger and Telescope Array observatories. The UHECR flux 
presents two main features, a hardening of the spectrum known as the ankle and a suppression at higher energies. The 
experimental data suggest that above the ankle the UHECRs flux is dominated by an extragalactic component of astrophysical 
origin. However, a minority component of exotic origin that dominates the flux beyond the suppression is still compatible 
with current data. Therefore, there exist the possibility that part of the UHECR flux originates from the decay of super-heavy 
dark matter particles clustered in the halos of the galaxies. In these scenarios the main contribution comes from the halo 
of our galaxy. In this article the possibility of identifying these scenarios in the context of the future very large exposure 
cosmic rays observatories is studied. It is worth mentioning that the contribution of the extragalactic halos located in 
the nearby universe is also included in these studies.}

\begin{document}
\maketitle
\flushbottom

\section{Introduction}
\label{sec:intro}

The nature of the ultra high energy cosmic rays (UHECRs, $E\geq 10^{18}$ eV) is still unknown. The main observables used to study 
its origin are the energy spectrum, the composition profile as a function of primary energy, and the distribution of their arrival
directions. 

The UHECR flux has been measured with good statistics by the Pierre Auger Observatory and Telescope Array. It presents two main 
features, a hardening at $\sim 10^{18.7}$ eV, known as the ankle and a suppression. This suppression is observed by Auger at 
$10^{(19.62 \pm 0.02)}$ eV and by Telescope Array at a larger energy, $10^{(19.78 \pm 0.06)}$ eV \cite{AugerTA:17}. Also, the Auger 
spectrum takes smaller values than the ones corresponding to Telescope Array. The discrepancies between the two observations can be 
diminished by shifting the energy scales of both experiments within their systematic uncertainties. However, some differences are 
still present in the suppression region \cite{AugerTA:17}.

The composition of the UHECRs is determined by comparing experimental data with air shower simulations, which makes use of high 
energy hadronic interaction models. These models present non negligible systematic uncertainties since the hadronic interactions
at the highest energies cannot be deduced from first principles. As a consequence, the composition determination is subject to
important systematic uncertainties. One of the most sensitive parameters to the nature of the primary is the atmospheric depth
of the maximum shower development, $X_{max}$. It can be obtained on an event by event basis from the data taken by the fluorescence
telescopes of Auger and Telescope Array. The mean value of $X_{max}$ obtained by Auger \cite{AugerXmax:14}, interpreted by using the 
updated versions of current high energy hadronic interaction models, shows that the composition is light from $\sim 10^{18}$ up to 
$\sim 10^{18.6}$ eV. From $\sim 10^{18.6}$ eV, the composition becomes progressively heavier for increasing values of the primary 
energy. This trend is consistent with the results obtained by using the standard deviation of the $X_{max}$ distribution 
\cite{AugerXmax:14}. On the other hand, the $X_{max}$ parameter reconstructed from the data taken by the fluorescence telescopes of 
Telescope Array is also compatible with a light composition at energies below the ankle, when it is interpreted by using the current 
hadronic interaction models \cite{TA:18}. It is worth mentioning that the $X_{max}$ distributions, as a function of primary energy, 
obtained by Auger and Telescope Array are compatible within systematic uncertainties \cite{Souza:17}. However, the presence of 
heavier primaries at energies above the ankle cannot be confirmed by the Telescope Array data due to the limited statistics of the 
event sample \cite{Souza:17}.  

The distribution of the arrival directions of the events with primary energies above $\sim 10^{18.9}$ eV detected by Auger
presents an anisotropy that can be described as a dipole of $\sim 6.5$\% amplitude \cite{Science:17}. The significance of this 
detection is larger than $5.2 \sigma$. The dipole direction is such that a scenario in which the flux is dominated by a galactic 
component is disfavored \cite{Science:17}. Regarding point source searches, Auger has found an indication of a correlation between
the arrival directions of the events of primary energy larger than $10^{19.6}$ eV and nearby starburst galaxies \cite{AugerStar:18}. 
The significance of this correlation is at $\sim 4\sigma$ level. Also the Auger data present an excess above $10^{19.76}$ eV
in the region of the radio galaxy Centaurus A \cite{AugerAnisICRC:17,AugerAnis:15}. However, the statistical significance of this 
excess is $\sim 3.1\sigma$. The Telescope Array Collaboration has also found an excess above $10^{19.75}$ eV in a direction of the 
sky which is contained in the supergalactic plane \cite{HotSpot:14,HotSpot:17}. The statistical significance of this excess is 
$\sim 3.4\sigma$.           

The experimental data suggest that the cosmic ray flux above the ankle is dominated by a component originated in extragalactic 
sources, most of those are possibly starburst galaxies. Besides, these sources accelerate not only protons but also heavier 
nuclei, assuming that current high energy hadronic interaction models do not present too large systematic uncertainties. However,
a minority component of another origin that could dominate the flux beyond the suppression is still compatible with the experimental 
data \cite{Alcantara:19}. 

The possibility that the by-products of the decay of unstable super-heavy dark matter (SHDM) particles can contribute to the UHECR 
flux has been studied extensively in the past (see for instance \cite{MedinaTanco:99,Aloisio:08,Kalashev:08,Aloisio:15,Kalshev:17,
Marzola:17}). In these models the dark matter is composed of supermassive particles produced gravitationally during inflation 
\cite{Kuzmin:98,Chung:98,Kuzmin:99a,Kuzmin:99b}. These particles would be clustered in the halo of the galaxies including ours. 
The spectrum of SHDM particles is expected to be dominated by gamma rays, protons and neutrinos. The upper limits to the gamma-ray 
flux obtained by Auger and the non detection of events above $10^{20.3}$ eV by Auger impose tight constraints to the flux 
corresponding to this hypothetical SHDM component. Therefore, to test the hypothesis of the existence of this component, observatories 
of very large exposure are required. In this article we study the possibility to identify this scenario in the context of the next 
generation UHECR observatories which will have a much larger exposure than current ones. In this study, besides considering the 
contribution of the galactic halo, the contribution of extragalactic halos located in the nearby universe is also included.

\section{Cosmic rays from galactic and extragalactic SHDM}

The rest mass and the decay time of the SHDM particles are free parameters in models in which a minority component of the UHECRs 
originate from the decay of these unstable particles. In the energy range of interest these parameters are constrained by cosmic ray
observations. In particular, the most restrictive constraints are imposed by the upper limits to the photon flux found by Auger
\cite{Kalashev:16} and also the non-observation by Auger of events above $10^{20.3}$ eV \cite{Alcantara:19}. This last analysis imposes 
more restrictive constraints than the ones based on the upper limits to the photon flux for scenarios in which $M_X>10^{23}$ eV. 
Therefore, the mass of the SHDM particles considered in this work is $M_X=10^{22.3}$ eV, for which only the constraints coming from the 
upper limits to the photon flux obtained by Auger are relevant.

Given the mass of the SHDM particles the decay time corresponding to the scenario for the largest SHDM cosmic ray flux, compatible with 
the upper limits to the photon fraction obtained by Auger, can be estimated from the predicted integral gamma-ray flux, which is given 
by,
\begin{eqnarray}
J_\gamma(>E)=&& \frac{1}{4 \pi\ M_X c^2\ \tau_X}\ \sum_{s=1}^N \int_E^\infty dE'\ \frac{dN_{\gamma,\, s}}{dE'}(E',D_s) \int_0^\infty d\xi %
\int_0^{2 \pi} d\alpha\ \int_0^{\pi} d\delta \cos\delta \times \nonumber \\ 
\label{IntG}
&&\rho_{X,\, s}(r(\xi,\alpha,\delta,\alpha_s,\delta_s)) \ \varepsilon(\delta), 
\end{eqnarray}
where $M_X$ is the rest mass of the SHDM particle, $\tau_X$ is its decay time, $c$ is the speed of light, $N$ is the number of 
dark matter halos considered, $\rho_{X,s}$ is the energy density of the $s-th$ dark matter halo, $r$ is the distance measured from 
the center of the halo to a given point in the space, $\alpha_s$ is the right ascension of the center of the $s-th$ halo, $\delta_s$ 
is the declination of the center of the $s-th$ halo, $\xi$ is the distance from the Earth in the direction defined by the angles 
$\alpha$ and $\delta$, $dN_{\gamma,s}/dE$ is the number of gamma rays per units of energy corresponding to a single decay including 
the effects of propagation, and $D_s$ is the comoving distance from the Earth to the center of the $s-th$ halo. Here 
$\varepsilon(\delta)$ is the relative exposure of Auger which fulfills the normalization condition,
\begin{equation}
\int_0^{\pi} d\delta \cos\delta \ \varepsilon(\delta) = 1.
\end{equation} 
An analytical expression for $\varepsilon(\delta)$ can be found in Ref.~\cite{Sommers:01}.

Gamma rays generated in SHDM decays can interact with low energy photons of the radiation field present in the universe during 
propagation. The relevant low energy photon backgrounds are the cosmic microwave background (CMB) and the radio background (RB). 
The main processes undergone by gamma rays are pair production ($\gamma+\gamma_b \rightarrow e^+ + e^-$) and double pair production 
($\gamma+\gamma_b \rightarrow e^+ + e^- + e^+ + e^-$).

Gamma rays that originate in our Galaxy are assumed to propagate freely due to the fact that the distances traveled by them, 
from generation to the Earth, are much smaller than their mean free path. However, the spectrum of gamma rays, which originated 
from SHDM decays in extragalactic halos are modified due to the interactions undergone by them during propagation. Therefore, the 
energy spectrum of the gamma rays at Earth takes the following form,
\begin{eqnarray}
\label{SpecG} 
&&\frac{dN_\gamma}{dE}(E,D)=\frac{dN_{\gamma}^0}{dE}(E) \ \ \ \ \textrm{Galactic gamma rays} \\
&&\frac{dN_\gamma}{dE}(E,D)=\exp\left[ -\frac{D}{\lambda_{\gamma \gamma}(E)}\right]\ \frac{dN_\gamma^0}{dE}(E) \ \ \ \ 
\textrm{Extragalactic gamma rays}, 
\label{SpecEG}
\end{eqnarray}   
where $dN_\gamma^0/dE$ is the energy distribution at decay, $\lambda_{\gamma \gamma}$ is the mean free path of gamma rays in 
the photon background, and $D$ is the distance from the center of the halo to the Earth. Note that Eq.~(\ref{SpecEG}) is valid 
for a non-expanding universe which in our case is a good approximation, since the extragalactic halos considered are at distances 
smaller than $\sim 140$ Mpc. 

The energy distributions $dN_{\gamma,\, p}^0/dE$ of the gamma rays and protons (the secondary particles considered in this work) 
generated from the decay of the SHDM particles are calculated by using the SHdecay program \cite{Barbot:04}.       

\begin{figure}[!h]
\centerline{\includegraphics[width=10cm]{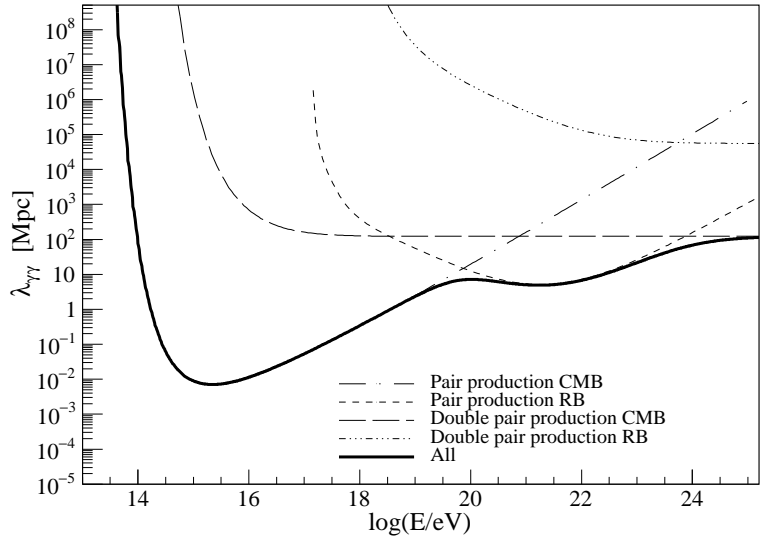}}
\caption{Mean free path of gamma rays in the CMB and RB as a function of the gamma ray energy. \label{GammaMFP}}
\end{figure}

Figure \ref{GammaMFP} shows the mean free path of gamma rays in the CMB and RB for the relevant processes. The radio background model 
used for the calculation is the one developed in Ref.~\cite{Protheroe:96}. The calculation is performed by using the tools developed 
for the package CRPropa 3, which are accessible at Ref.~\cite{CRPropa3Data}. It can be seen that for energies above $10^{19}$ eV the 
total mean free path is larger than 1 Mpc, this is the reason that supports the assumption that gamma rays originated in the halo of 
our Galaxy propagate freely. It can also be seen that below $\sim 10^{19.5}$ eV the mean free path is dominated by pair production in 
the CMB, from $\sim 10^{19.5}$ eV to $\sim 10^{23.5}$ eV the relevant process is also pair production but in the RB, and finally above 
$\sim 10^{23.5}$ eV the dominant process is the double pair production in the CMB. 

The Burkert dark matter profile \cite{Burkert:95} is considered in this work. It is given by,
\begin{equation}
\rho_X(r)=\frac{\rho_B}{\left(1+\frac{r}{r_B}\right) \left( 1+\left(\frac{r}{r_B}\right)^2 \right)},
\end{equation}
where $\rho_B$ and $r_B$ depend on the halo under consideration. For the Milky Way the parameters are $\rho_B=1.187$ GeV cm$^{-3}$
and $r_B=10$ kpc \cite{Nesti:13}. The dark matter halos considered are the ones of the DMCat catalog \cite{Lisanti:18a,Lisanti:18b}, 
which is based on the galaxy group catalogs of Refs.~\cite{Tully:15,Kourkchi:17}. In that catalog there are 17021 halos with 
comoving distances smaller than $\sim 140$ Mpc. The parameters of the Burkert profile corresponding to these extragalactic 
halos are can obtained from the catalog. 

\begin{figure}[!h]
\centerline{\includegraphics[width=10cm]{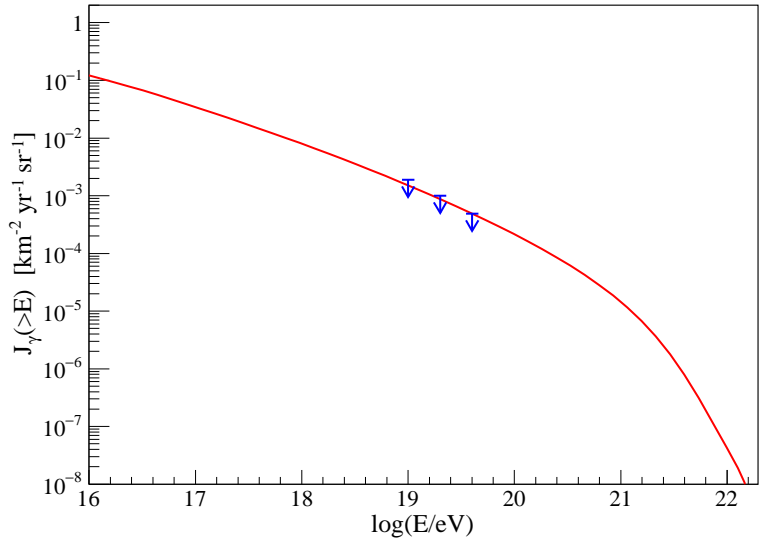}}
\caption{Integral gamma-ray flux as a function of the logarithm of the energy. The galactic and extragalactic (halos in the 
DMCat catalog) contributions are included. The arrows correspond to the 95\% CL upper limits obtaind by Auger \cite{PhLimits:15}.
\label{IntGammaAuger}}
\end{figure}

Figure \ref{IntGammaAuger} shows the integral gamma-ray flux obtained by using Eq.~(\ref{IntG}) for $\tau_X = 5.4\times 10^{22}$ yr.
This value for the decay time corresponds to the largest integral gamma-ray flux compatible with the Auger upper limits 
\cite{PhLimits:15}, which are also shown in the figure. The contributions of our galaxy and the one corresponding to the halos in the 
DMCat catalog are included.

The cosmic ray flux originated in SHDM decays for an observatory with uniform exposure is given by, 
\begin{eqnarray}
J_i(E)=&& \frac{1}{(4 \pi)^2 \ M_X c^2\ \tau_X}\ \sum_{s=1}^N \frac{dN_{i,\, s}}{dE}(E,D_s) \int_0^\infty d\xi %
\int_0^{2 \pi} d\alpha\ \int_0^{\pi} d\delta \cos\delta \times \nonumber \\ 
\label{IntI}
&&\rho_{X,\, s}(r(\xi,\alpha,\delta,\alpha_s,\delta_s)), 
\end{eqnarray}
where $i\in\{p,\gamma\}$, $dN_{\gamma,\, s}/dE$ is given by Eq.~(\ref{SpecG}) for the galactic halo and Eq.~(\ref{SpecEG}) for
the extragalctic halos, and
\begin{eqnarray}
\label{SpecPrG} 
&&\frac{dN_p}{dE}(E,D)=\frac{dN_{p}^0}{dE}(E) \ \ \ \ \textrm{Galactic protons} \\
&&\frac{dN_p}{dE}(E,D)=\int_0^\infty dE' P(E|E',D)\ \frac{dN_p^0}{dE'}(E') \ \ \ \ 
\textrm{Extragalactic protons}. 
\label{SpecPrEG}
\end{eqnarray}   
Here $dN_p^0/dE$ is the proton energy distribution at decay and $P(E|E',D)$ is the energy distribution of a proton at Earth 
with energy $E$ injected at a comoving distance $D$ with energy $E'$. 

As well as gamma rays, extragalactic protons can undergo interactions with the low energy photon backgrounds during propagation 
through the Universe. The main processes are pair production ($p+\gamma_b \rightarrow p+e^+ + e^-$) and photopion production 
($p+\gamma_b \rightarrow N+\pi s$, where $N$ corresponds to nucleons). The distribution function $P(E|E',D)$ is obtained from 
simulations performed by using the CRPropa 3 package \cite{CRPropa3:16}. In this program all relevant processes are taken into 
account including also the interactions of the ultra high energy protons with the low energy photons of the extragalactic 
background light (see Ref.~\cite{CRPropa3:16} for details). 

\begin{figure}[!h]
\centerline{\includegraphics[width=11cm]{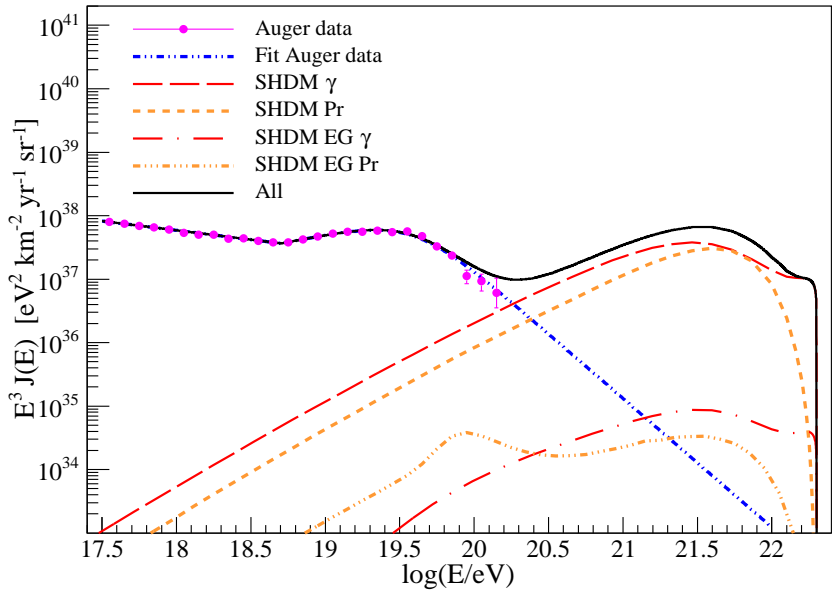}}
\caption{Ultra high energy cosmic ray flux as a function of the logarithm of the primary energy. The data points correspond
to the Auger measurements \cite{AugerSpec:17}. Double dotted-dashed line corresponds to a fit of the Auger data. Dashed lines 
correspond to the proton and gamma-ray components originated from SHDM decays and the triple dotted-dashed and dotted-dashed 
lines correspond to the proton and gamma-ray contribution from the extragalactic halos of the DMCat catalog, respectively. 
Solid line corresponds to the total contribution. 
\label{FluxAuger}}
\end{figure}

Figure \ref{FluxAuger} shows the energy spectrum observed by Auger \cite{AugerSpec:17} fitted with the function defined in 
Ref.~\cite{AugerSpec:17} (see appendix \ref{FluxAstro}). In the scenario considered in this work it is assumed that this 
component is of astrophysical origin. The Auger data are compatible with a SHDM component which starts to dominate the flux 
above $10^{20}$ eV. The SHDM contribution, which corresponds to the considered scenario in which $M_X=10^{22.3}$ eV and 
$\tau_X=5.4\times10^{22}$ yr is also shown in the figure. As before, the contributions of the galactic halo and the extragalactic 
halos from the DMCat catalog are included. From the figure it can be seen that the contribution from the extragalactic halos is a 
small fraction of the total SHDM flux. The propagation effects on the proton and gamma-ray components are also evident, in particular 
the proton component presents a pile-up originated from the photopion production process \cite{Bere:88}.          

\begin{figure}[!h]
\centerline{\includegraphics[width=7.7cm]{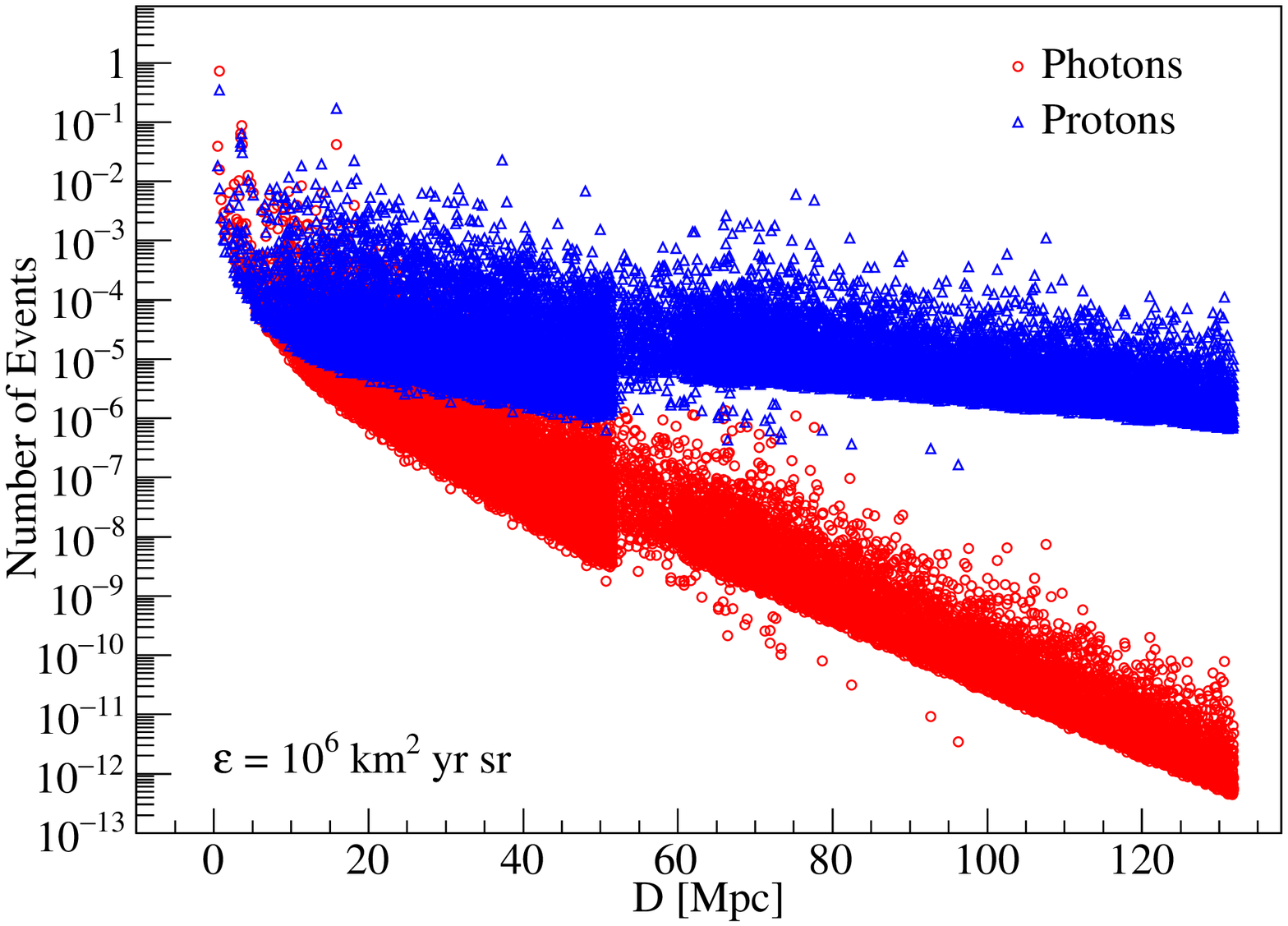}
\includegraphics[width=7.7cm]{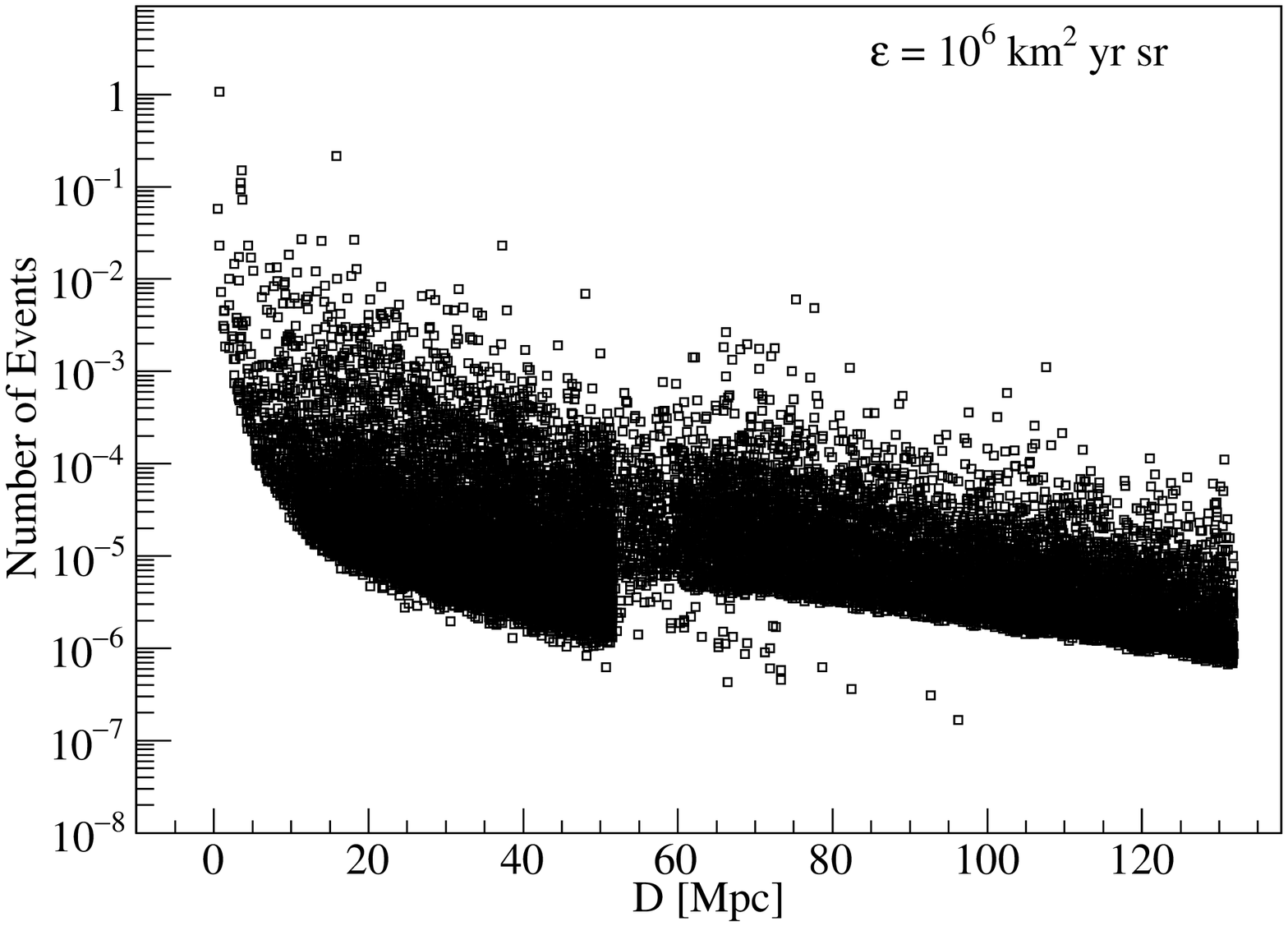}}
\caption{Expected number of events originated from SHDM decays, corresponding to each halo in the DMCat catalog, as a function 
of the comoving distance $D$ for exposure $\mathcal{E} = 10^6$ km$^2$ yr sr and energy above $10^{20}$ eV. Left panel: expected 
number of events for protons and gamma rays. Right panel: total number of events irrespective of the type. \label{NEvents}}
\end{figure}

One of the most important characteristic of next generation space-based UHECR observatories, like JEM-EUSO \cite{JEMEUSO} 
and POEMMA \cite{POEMMA}, is its very large exposure. Fig.~\ref{NEvents} shows the expected number of events originated from 
SHDM decays, corresponding to each halo in the DMCat catalog, as a function of the comoving distance for a constant exposure 
of $\mathcal{E} = 10^6$ km$^2$ yr sr above $10^{20}$ eV \cite{POEMMA}. The left panel of the figure shows the separated proton 
and gamma ray contributions and the right panel shows the sum of these two. Note that the number of events corresponding to 
proton (photon) primaries is given by the integral of the corresponding term in Eq.~(\ref{IntI}) above $10^{20}$ eV, with $i=p$ 
($i=\gamma$), multiplied by the exposure $\mathcal{E}$. From the left panel of the figure it can be seen that the number of 
events corresponding to gamma rays decreases much faster with the comoving distance than the one corresponding to protons. This 
is due to the fact that the gamma rays undergoing interactions are removed from the flux but protons, or more precisely nucleons, 
lose a fraction of their energy in each interaction but they still contribute to the flux. From the right panel of the figure it
can be seen that there is only one halo, located close to the Earth, for which the expected number of events is of order one. 
For the rest of the halos is smaller than $\sim 0.3$. The halo that most contributes to the number of events corresponds to the 
Andromeda galaxy, also known as M31. Note that in the two plots of Fig.~\ref{NEvents} two different regions can be identified,
which are separated at a comoving distance of $\sim 55$ Mpc. These two regions correspond to the two different galaxy catalogs 
used to build the DMCat catalog \cite{Lisanti:18a}.                  

Although the contribution from a given extragalactic halo is much smaller than the one corresponding to the galactic halo, it 
can be important due to the fact that the gamma rays and protons originating in such halo come from a narrow region of the sky. 
Therefore, in that region the contribution of the extragalactic halo can be more important than the one corresponding to our 
galaxy, specially in regions far from the galactic center where the galactic contribution decreases considerably. In order
to study this possibility the angular distribution of gamma rays and proton is required. The flux from a given halo, $s$, is 
given by,   
\begin{eqnarray}
J_{s,i}(E,\theta) = && \frac{\textrm{sr}^{-1}}{4 \pi\ M_X c^2\ \tau_X}\ \frac{dN_{s,i}}{dE}\ \left[ 2\ \Theta\left(\frac{\pi}{2}-
\theta\right) \int_{D_s \sin\theta}^{D_s} dr \frac{r\ \rho_{X,\, s}(r)}{\sqrt{r^2-D_s^2 \sin^2(\theta)}} + \right. \nonumber \\
&&\left. \int_{D_s}^{\infty} dr \frac{r\ \rho_{X,\, s}(r)}{\sqrt{r^2-D_s^2 \sin^2(\theta)}} \right],
\label{JSHDMAng}
\end{eqnarray}
where $\Theta(x)$ is the Heaviside function (i.e.~$\Theta(x)=1$ for $x \geq 0$ and $\Theta(x)=0$ otherwise) and $\theta \in [0,\pi]$ 
is the angle between the direction of the center of the halo and the direction of observation. Note that, for the Burkert dark matter 
profile, the integral in Eq.~(\ref{JSHDMAng}) can be done analytically (see appendix \ref{AngDist} for details).

Figure \ref{AngDistPlot} shows the angular distribution calculated from Eq.~(\ref{JSHDMAng}) (see appendix \ref{AngDist}), 
normalized to its value at $\theta=0$, for the Milky Way (left panel) and Andromeda (right panel). It can be seen from the figure 
that even though the distribution of Andromeda is much narrower than the one corresponding to the Milky Way, it has a non negligible 
angular width, which is larger than $4^\circ$.  
\begin{figure}[!h]
\centerline{\includegraphics[width=7.7cm]{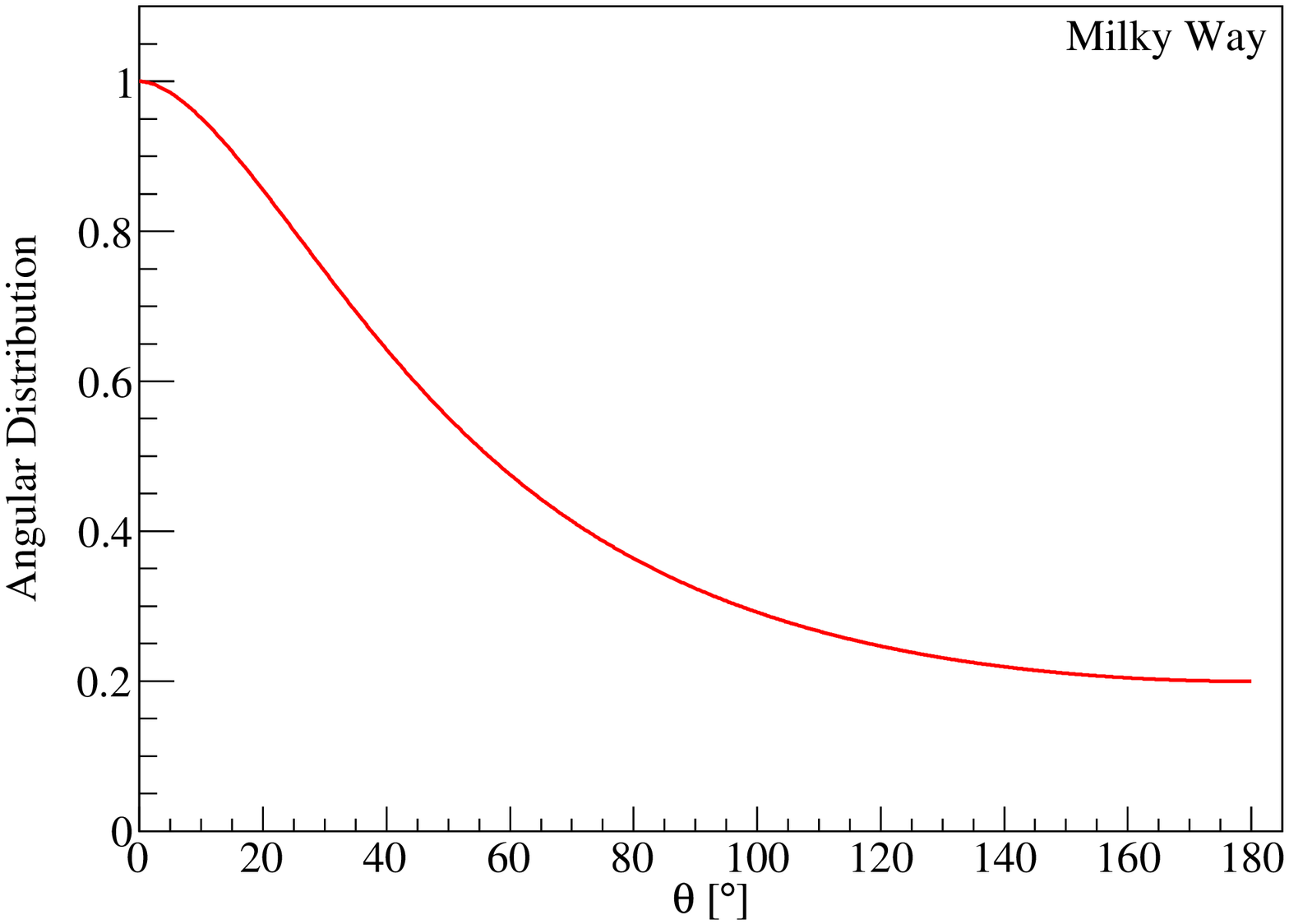}
\includegraphics[width=7.7cm]{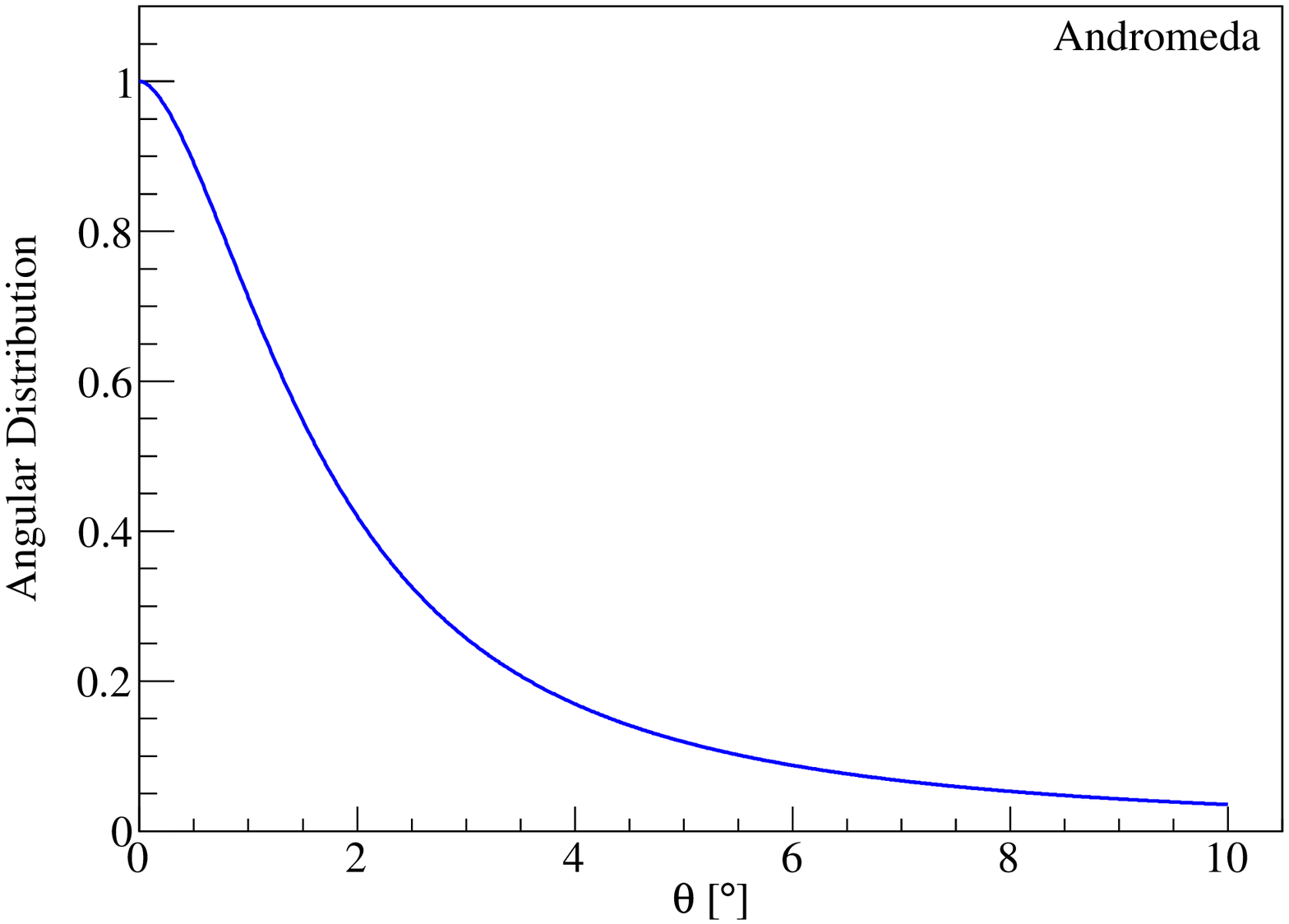}}
\caption{Angular distribution, normalized to its value at $\theta=0$, as a function of $\theta$ for the Milky Way (left panel)
and for Andromeda (right panel). \label{AngDistPlot}}
\end{figure}

The HEALPix library \cite{Healpix} is used to study the contribution of the different UHECRs sources in a given region of the 
sky. It is assumed that the arrival directions of the cosmic rays of astrophysical origin are uniformly distributed. Given a
pixelization of the sphere the average number of events with primary energies above $E_{min}$ and arrival directions contained 
in the $j-th$ pixel is $\langle n_j \rangle(E_{min})=\langle n_j \rangle_{Astro}(E_{min}) + \langle n_j \rangle_{SHDM}(E_{min})$, 
where
\begin{eqnarray}
\label{Nastro}
&&\langle n_j \rangle_{Astro}(E_{min}) = \mathcal{E} \ \frac{\Omega_j}{4 \pi} \ \int_{E_{min}}^\infty dE \ J_{Astro} (E), \\
\label{Nshdm}
&&\langle n_j \rangle_{SHDM}(E_{min}) = \mathcal{E} \ \frac{1}{4 \pi} \ \int_{E_{min}}^\infty dE \int_{\Omega_j} dl\ db \cos b 
\sum_{s=1}^N \sum_{i=\gamma,p} 
J_{s,i} (E,l,b).
\end{eqnarray}    
Here $\langle n_j \rangle_{Astro}(E_{min})$ corresponds to the average number of events of astrophysical origin, $\Omega_j$ is 
the solid angle subtended by the $j-th$ pixel, $J_{Astro} (E)$ is the flux of astrophysical origin (see appendix \ref{FluxAstro}), 
$\langle n_j\rangle_{SHDM}(E_{min})$ corresponds to the average number of events for the SHDM component, $l$ and $b$ are the 
galactic longitude and the galactic latitude, respectively, and $J_{s,i} (E,l,b)$ is given by Eq.~(\ref{JSHDMAng}) but written 
as a function of the galactic coordinates. Note that an observatory with uniform exposure is considered.  

\begin{figure}[!h]
\centerline{\includegraphics[width=7.7cm]{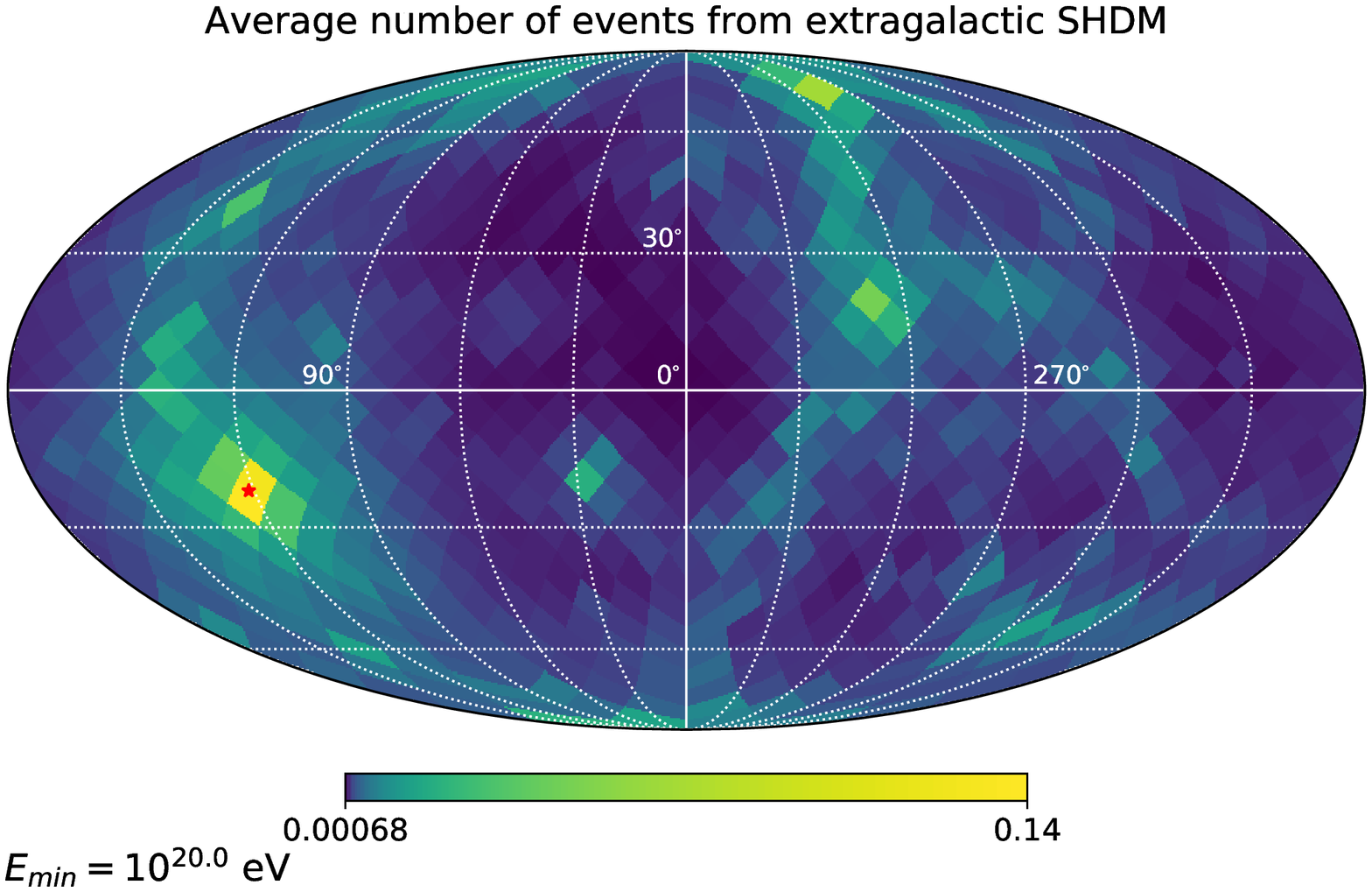}
\includegraphics[width=7.7cm]{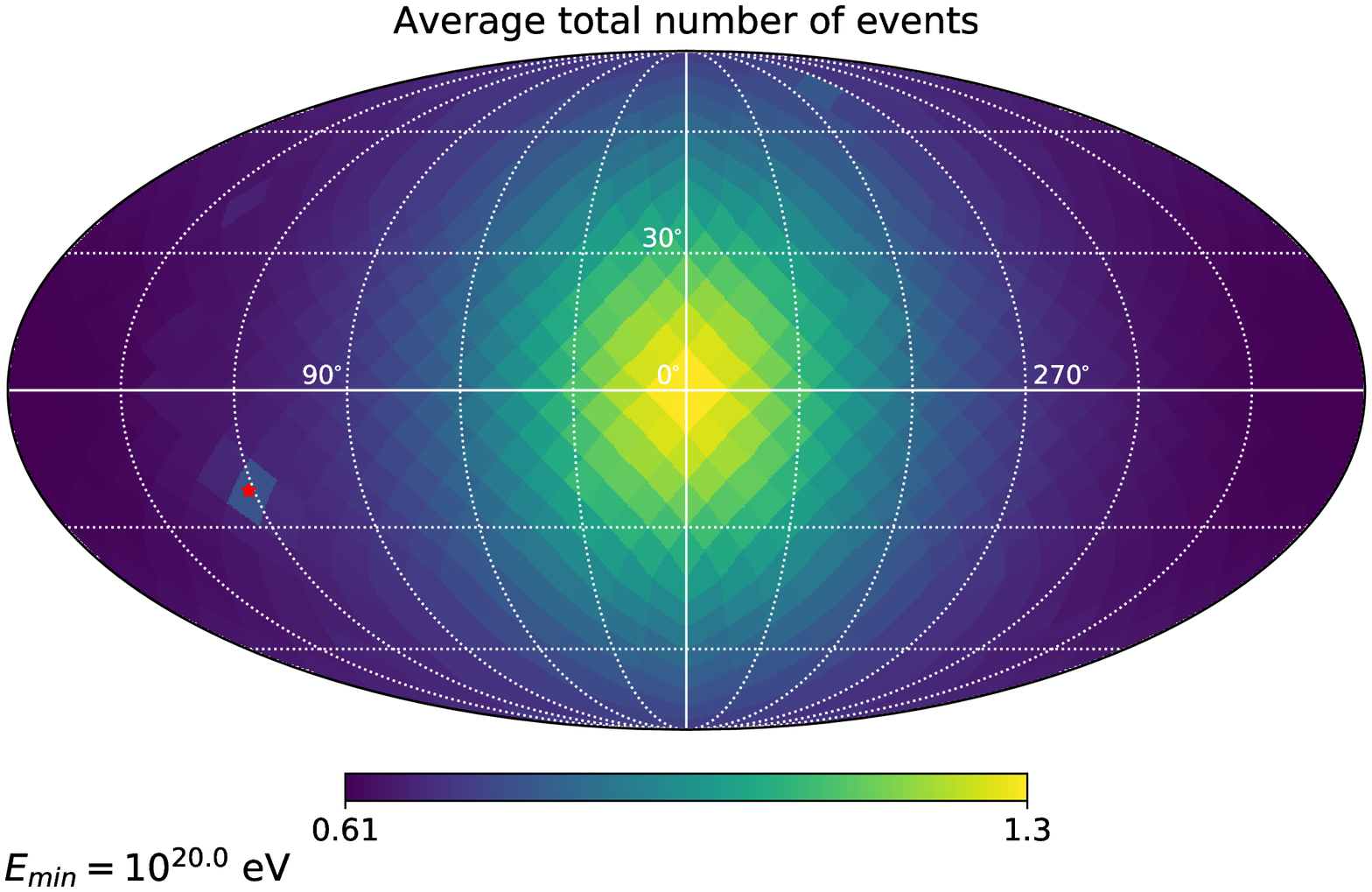}}
\caption{Left panel: Average number of events for the extragalactic halos from the DMCat catalog, a logarithmic scale color 
is considered in this case. Right panel: Average total number of events including the contributions from the decay of SHDM 
in the galactic halo, SHDM in the DMCat catalog halos, and the astrophysical component. Here $E_{min}=10^{20}$ eV and 
$\mathcal{E} = 10^6$ km$^2$ yr sr. The red star corresponds to the position of Andromeda. \label{SkyMapNev}}
\end{figure}

A pixelization of the sphere with 768 pixels is considered (corresponding to $N_{side} = 8$ \cite{Healpix}). In this pixelization
each pixel has an angular radius of the order of $4^\circ$. It is worth mentioning that the reconstruction uncertainties are not 
included in the calculation because at these energies the angular resolution is in general smaller than $1^\circ$ \cite{POEMMA}, which
is much smaller than the pixel radius. The angular integrals in Eq.~(\ref{Nshdm}) are performed by using the Monte Carlo technique. The 
left panel of Fig.~\ref{SkyMapNev} shows the average number of events expected in each pixel for the extragalactic halos from the DMCat
catalog, $E_{min}=10^{20}$ eV, and $\mathcal{E} = 10^6$ km$^2$ yr sr. Note that the scale color is logarithmic. From the figure it can 
be seen that the region in the sky for which the average number of events is larger corresponds to the surroundings of Andromeda. However, 
a larger exposure is required in order to increase the probability of observing at least one event in the pixels of the surroundings of 
Andromeda. The right panel of the figure shows the average total number of events, which includes the contributions from the decay of 
SHDM in the galactic halo, SHDM in the DMCat catalog halos, and the astrophysical component. It can be seen that the galactic halo 
dominates the average total number of events. However, Andromeda is placed in a region far from the galactic center, then its contribution 
can be significant, provided the total exposure is larger.        

The probability to observe a given number of events in a given pixel follows the Poisson distribution. This probability strongly 
depends on the scenario considered. In particular, for a given value of the exposure, the probability to observe at least one 
event in the pixels corresponding to the galactic center region is larger for the case in which the contribution from the 
SHDM decays is non negligible. The probability to observe at least one event in the $j-th$ pixel is given by,
\begin{equation}
\label{Proba}
P(n_j \geq 1 | E_{min})= 1-\exp(-\mu_j),
\end{equation}  
where $\mu_j = \langle n_j \rangle(E_{min})$ or $\mu_j = \langle n_j \rangle_{Astro}(E_{min})$ for the cases in which the SHDM 
contribution is non negligible and negligible, respectively. Therefore, the exposure required to measure at least one event 
in the $j-th$ pixel with probability $p_0$ is given by,
\begin{equation}
\widetilde{\mathcal{E}}_j(p_0)=-\frac{1}{\mu_j}\ \ln(1-p_0).
\end{equation}

\begin{figure}[!h]
\centerline{\includegraphics[width=7.7cm]{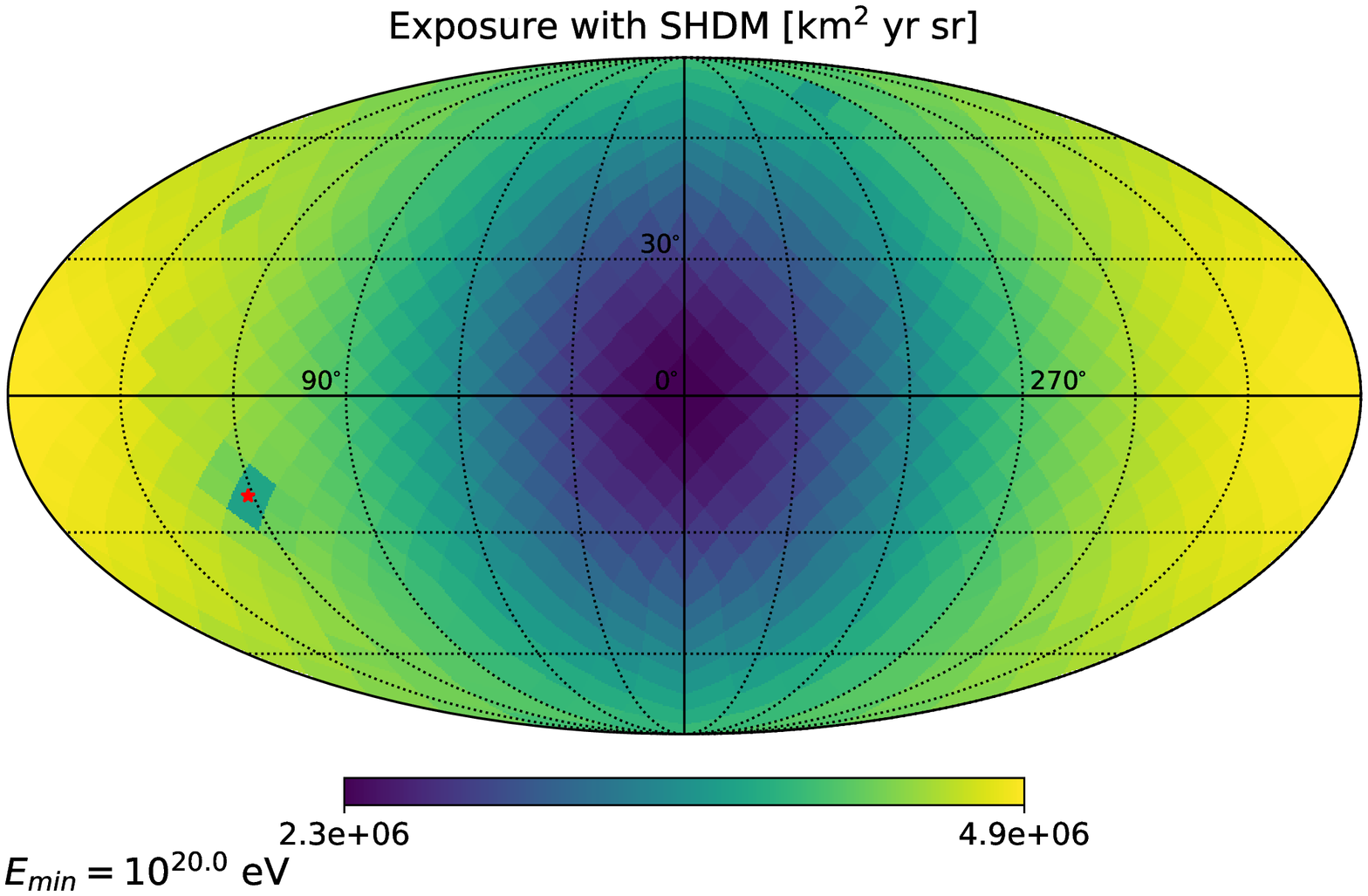}
\includegraphics[width=7.7cm]{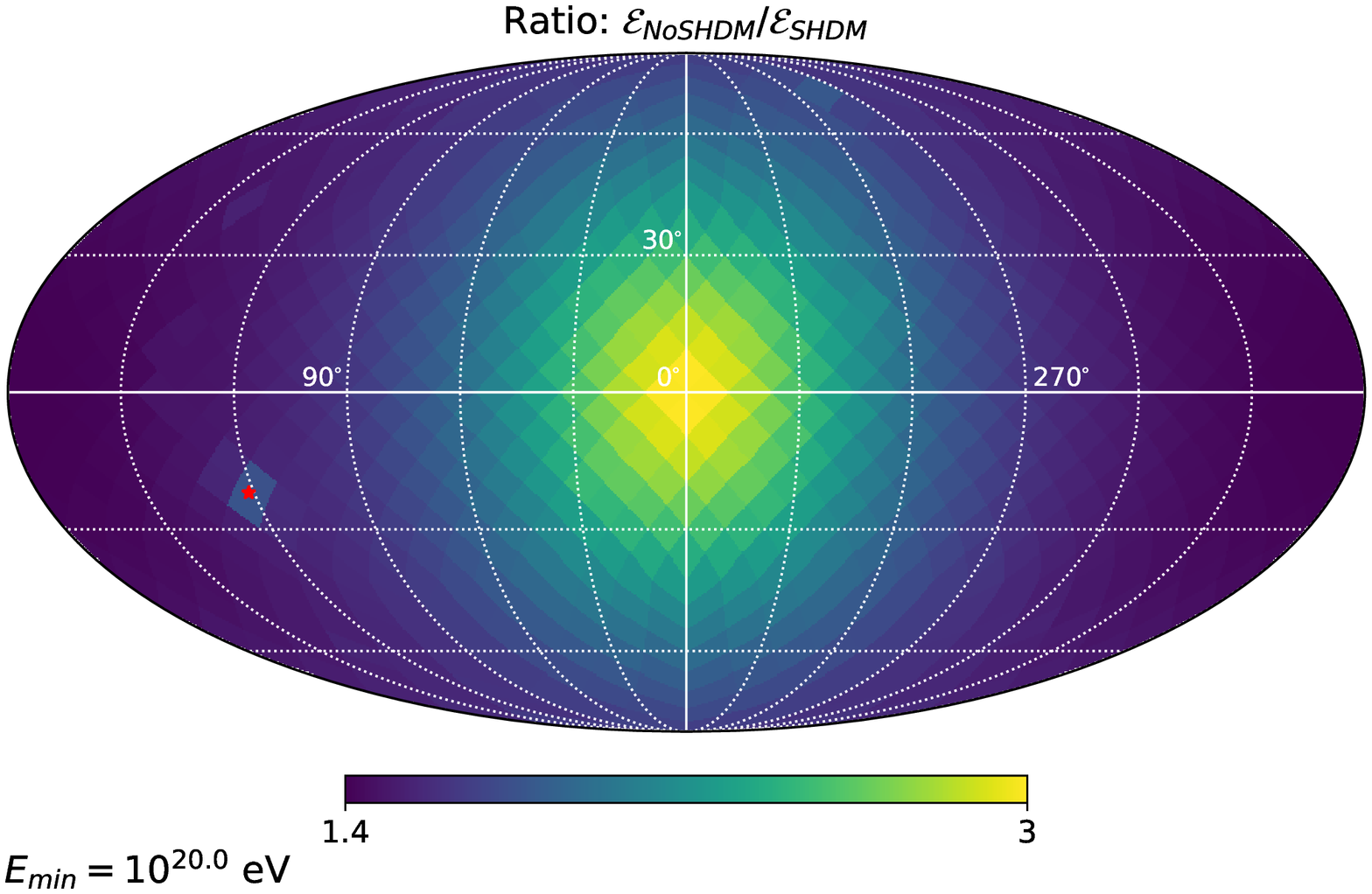}}
\caption{Left panel: Exposure required to observe at least one event in each pixel with 0.95 probability including all 
contributions considered. Right panel: Ratio between the exposure required to observe at least one event in each pixel 
with 0.95 probability without and with the inclusion of a non negligible SHDM component. The minimum energy is 
$E_{min}=10^{20}$ eV and the red star corresponds to the position of Andromeda. 
\label{SkyMapExp}}
\end{figure}

The left panel of Fig.~\ref{SkyMapExp} shows the exposure required to observe at least one event in each pixel with 0.95 
probability, i.e.~$\widetilde{\mathcal{E}}_j(0.95)$ for $E_{min}=10^{20}$ eV and for the case in which the contribution from 
SHDM is non negligible. It can be seen that $\widetilde{\mathcal{E}}_j(0.95)$ corresponding to the pixels in the center 
of the galaxy is more than two times smaller than the one corresponding to the regions far from the galactic center. In the 
case of the Andromeda region $\widetilde{\mathcal{E}}_j(0.95)$ has to be $\sim 1.3$ times smaller compared to regions far from 
the galactic center. The right panel of Fig.~\ref{SkyMapExp} shows the ratio between $\widetilde{\mathcal{E}}_j(0.95)$ for the 
scenario with a negligible SHDM contribution and for the one with a non negligible SHDM contribution. It can be seen that in 
the galactic center region $\widetilde{\mathcal{E}}_j(0.95)$ for the scenario with a negligible SHDM contribution has to be 
three times larger than the one corresponding to the case with a non negligible SHDM component. In the Andromeda region this 
ratio is $\sim 1.8$.     

For $E_{min}=10^{20.3}$ eV $\widetilde{\mathcal{E}}_j(0.95)$ increases (see the left panel of Fig.~\ref{SkyMapExp20.3} 
of appendix \ref{RExpo}). In this case $\widetilde{\mathcal{E}}_j(0.95)$ ranges from $6.3\times 10^6$ km$^2$ yr sr to 
$2.5\times 10^7$ km$^2$ yr sr. Also the relative importance of the SHDM component of the flux increases for increasing 
values of the minimum energy, making the differences between the two scenarios considered more important. In particular, 
in the galactic center region, $\widetilde{\mathcal{E}}_j(0.95)$ for the case with a negligible SHDM contribution has 
to be 17 times larger than the one corresponding to the case with a non negligible SHDM component. Also, in the Andromeda 
region it has to be $\sim 7.5$ times larger (see the right panel of Fig.~\ref{SkyMapExp20.3} of appendix \ref{RExpo}). 
Therefore, the observation of at least one event in one of those pixels for a given exposure can be used to discriminate 
between these two different scenarios.

The probability to observe at least one event in a given set of pixels is given by,
\begin{equation}
P_{set}(n \geq 1 | E_{min}, \mathcal{E})= 1-\exp\left[-\sum_{j\in S_p} \mu_j(E_{min}, \mathcal{E}) \right],
\end{equation}
where $S_p$ is the set of pixels considered. Fig.~\ref{Proba} shows the probability to observe at least one event in two different 
sets of pixels as a function of the exposure for the two scenarios considered and for $E_{min}=10^{20}$ eV and $E_{min}=10^{20.3}$ eV. 
The two sets of pixels considered are: \emph{i}) the four pixels closer to the galactic center, $S_{GC}$, and \emph{ii}) the two 
hottest pixels in the surroundings of Andromeda (see Fig.~\ref{SkyMapNev}), $S_{M31}$. From the figure it can be seen that the 
probability reaches one for smaller values of the exposure in the scenarios with a non negligible SHDM contribution. Also, the
probability reaches one for smaller values of the exposure in the case of $S_{GC}$, as expected.   
\begin{figure}[!h]
\centerline{\includegraphics[width=7.7cm]{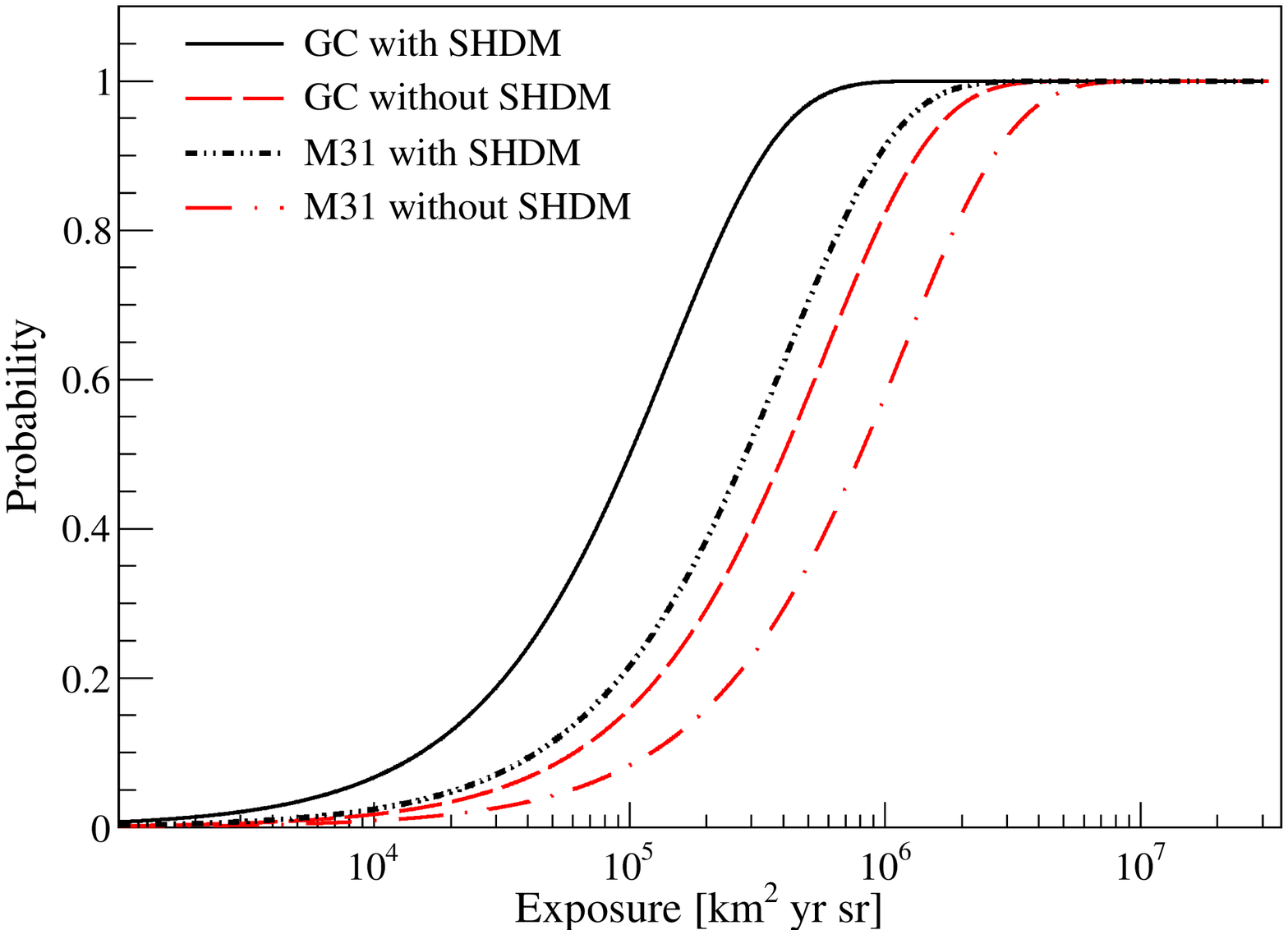}
\includegraphics[width=7.7cm]{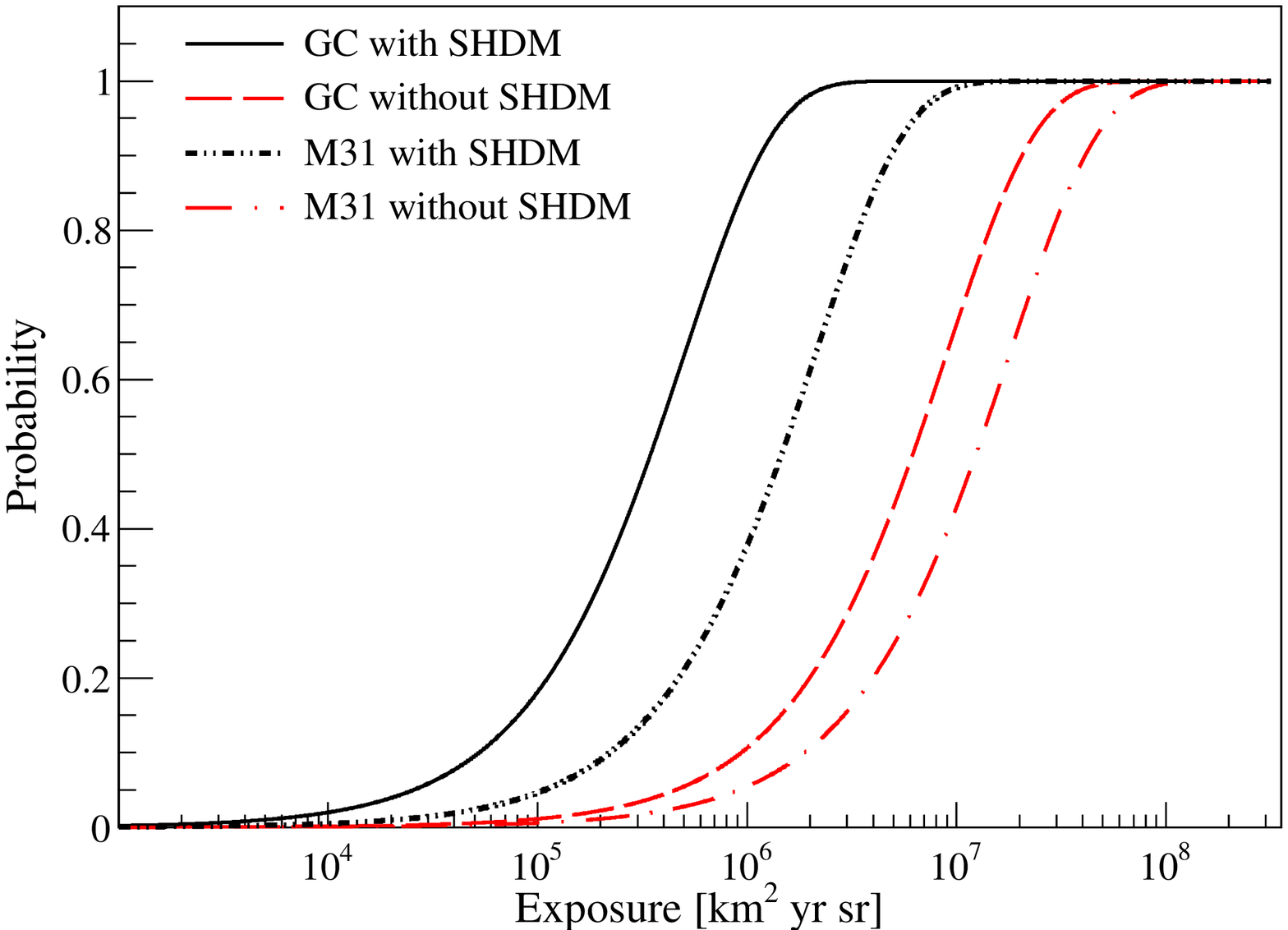}}
\caption{Probability to observe at least one event in a given set of pixels (see text for details) as a function of the 
exposure for $E_{min}=10^{20}$ eV (left panel) and $E_{min}=10^{20.3}$ eV (right panel). \label{Proba}}
\end{figure}

For a given set of pixels let us consider the exposure for which the probability to observe at least one event in the scenario without 
a component originated from SHDM is $0.1$, which is denoted as $\mathcal{E}_{10}$. Therefore, if for this value of the exposure reached
by a given observatory at least one event is observed in this set of pixels, the null hypothesis that states that the UHECR flux is the 
one corresponding to the astrophysical origin is rejected at $90\%$ confidence level (CL). The probability to observe at least one event 
for the exposure value $\mathcal{E}_{10}$ in a given set of pixels but for the model that includes a SHDM contribution considered before
gives the probability to reject the null hypothesis. This probability will be denoted as $P_{rej}$. Table \ref{P10E10} shows the values 
of $\mathcal{E}_{10}$ and $P_{rej}$ for $E_{min}=10^{20}$ eV and $E_{min}=10^{20.3}$ eV and for the two sets of pixels considered. From 
the table it can be seen that for $E_{min}=10^{20}$ eV, $P_{rej}$ is smaller than 0.5 for the two sets of pixels considered (0.34 and 
0.26 for $S_{GC}$ and $S_{M31}$, respectively). Note that the Auger exposure at present is approximately $8 \times 10^4$ km$^2$ yr sr 
\cite{Verzi:19} and it can reach values of order of $2 \times 10^5 $ km$^2$ yr sr at the end of its life \cite{Batista:18}. Therefore, 
with Auger data it will be possible to performed the proposed test for $E_{min}=10^{20}$ eV. For $E_{min}=10^{20.3}$ eV $P_{rej}$ is 
larger than 0.5, i.e.~0.85 and 0.59 for $S_{GC}$ and $S_{M31}$, respectively. The set $S_{M31}$ requires a larger exposure but it can 
be used to test independently the null hypothesis. 
\begin{table}[ht]
\begin{center}
\begin{tabular}{|c|c|c|c|c|} 
\hline
  $E_{min}$ [eV] & $\mathcal{E}_{10}$ [km$^2$ yr sr] for $S_{GC}$ &  $P_{rej}$ for $S_{GC}$  &%
  $\mathcal{E}_{10}$ [km$^2$ yr sr] for $S_{M31}$ &  $P_{rej}$ for $S_{M31}$ \\[0.5ex] 
 \hline
  $10^{20}$  & $6.1\times 10^4$   &  0.34  & $1.2\times 10^5$ & 0.26 \\[0.5ex] 
 \hline
 $10^{20.3}$ & $9.4\times 10^5$   &  0.85  & $1.9\times 10^6$ & 0.59 \\[0.5ex] 
 \hline
\end{tabular}
\caption{$\mathcal{E}_{10}$ and $P_{rej}$ for the two sets of pixels considered and for $E_{min}=10^{20}$ eV and $E_{min}=10^{20.3}$ eV.}
\label{P10E10}
\end{center}
\end{table}

Although for $E_{min}=10^{20.3}$ eV larger values of the exposure are required for the probability to saturate, it allows to 
reject the null hypothesis by using both set of pixels considered. The reason for this behavior has to do with the fact that 
in this energy range the contribution of the astrophysical component decreases much faster (the flux goes as $\sim E^{-5}$) 
than the one corresponding to the SHDM component (see Fig.~\ref{FluxAuger}) and then the component originated from SHDM decay 
becomes more important.

\section{Conclusions}

In this article the possibility to identify a scenario in which there is a non-negligible but minority component, originated 
from the decay of SHDM particles, that dominates the UHECR flux beyond the suppression has been studied. Due to the expected 
small flux of UHECRs originated from the decay of SHDM particles these studies have been done in the context of the next 
generation UHECR observatories which are planed to have larger exposures compared with current ones. Besides the contribution 
from the galactic halo, the contribution from extragalactic halos has also been considered.

The scenario in which the SHDM particles have a mass of $10^{22.3}$ eV and a decay time of $\tau_X = 5.4\times 10^{22}$ yr has
been considered. The values of these two parameters are compatible with current constraints. For this scenario it has been found 
that the halo of the Andromeda galaxy is the one that most contributes to the SHDM extragalactic component. For a uniform exposure 
of $10^6$ km$^2$ yr sr the mean number of events expected from Andromeda, above $10^{20}$ eV, is of order one. The null hypothesis 
which states that the UHECR flux is composed by a uniform flux of astrophysical origin has a $\sim 85\%$ probability to be rejected
considering a set of pixels in the region of the sky close to the galactic center and for primary energies above $10^{20.3}$ eV. 
In this case the exposure required is $\sim 9.4\times 10^5$ km$^2$ yr sr. For larger values of the exposure, 
i.e.~$1.9\times 10^6$ km$^2$ yr sr, the null hypothesis has $\sim 59\%$ probability to be rejected considering the hottest pixels 
in the surroundings of Andromeda and considering also the same energy range. This can be used as an independent test. Therefore, 
the next generation UHECR observatories that reach exposures of the order of $10^6$ km$^2$ yr sr will be able to identify or even 
constraint the scenarios in which there is a minority component originated from the decay of SHDM particles.

\appendix
\section{Cosmic rays of astrophysical origin}
\label{FluxAstro}

The fitting function of Ref.~\cite{AugerSpec:17} is used to describe the component of astrophysical origin, which is given by,
\begin{equation}
J(E) = J_a \left\{
\begin{array}{ll}
\left( \mathop{\displaystyle \frac{E}{E_a} } \right)^{-\gamma_1} & E \leq E_a \\[0.4cm]
\left( \mathop{\displaystyle \frac{E}{E_a} } \right)^{-\gamma_2} 
\frac{1+\left( \frac{E_a}{E_s} \right)^{\Delta \gamma}}%
{1+\left(\frac{E}{E_s} \right)^{\Delta \gamma}} & E > E_a 
\end{array}
\right.,
\label{JCR}
\end{equation}
where $J_a$ is a normalization constant, $E_a=5.08\times 10^{18}$ eV, $E_s=3.9\times 10^{19}$ eV, $\gamma_1=3.293$,
$\gamma_2=2.53$, and $\Delta \gamma = 2.5$.

The integral flux for $E>E_a$, which is used in the calculations, is given by,
\begin{eqnarray}
J(>E)=&& J_a \ \left[1+\left( \frac{E_a}{E_s} \right)^{\Delta \gamma} \right] \left( \frac{E_a}{E_s} \right)^\gamma  
\left( \frac{E}{E_s} \right)^{1-\gamma-\Delta \gamma} \frac{E_s}{\gamma+\Delta \gamma-1} \times \nonumber \\[0.4cm]
&& _2F_1\left(1,\frac{\gamma+\Delta \gamma-1}{\Delta \gamma},\frac{\gamma+2 \Delta \gamma-1}{\Delta \gamma}, 
-\left(\frac{E_s}{E}\right)^{\Delta \gamma}   \right),
\end{eqnarray} 
where $_2F_1(a,b,c,z)$ is the hypergeometric function.

\section{Angular distribution}
\label{AngDist}

The integral in Eq.~(\ref{JSHDMAng}) can be performed analytically for the case of the Burkert dark matter profile. It can be 
expressed as,  
\begin{equation}
\label{JSHDMAngI}
J(E,\theta) = \frac{\textrm{sr}^{-1}}{4 \pi\ M_X c^2\ \tau_X}\ \frac{dN}{dE} \times
\left\{
\begin{array}{ll}
I_1(\theta,D) + I_2(\theta,D) & \ \ \ 0 \leq \theta \leq \frac{\pi}{2} \\[0.4cm]
I_2(\theta,D) - I_1(\theta,D) & \ \ \ \frac{\pi}{2} < \theta \leq \frac{\pi}{2}  
\end{array}
\right.,
\end{equation}
where subscripts $i$ and $s$ are omitted for clarity and,
\begin{eqnarray}
I_1(\theta,D) = && \frac{\rho_{B}\ r_{B}}{2} \left[ \frac{r_{B}}{\sqrt{D^2 \sin^2\theta + r_B^2}} 
\left( \arctan\left[ \frac{D | \cos\theta |}{\sqrt{D^2 \sin^2\theta + r_B^2}} \right] + \right. \right. \nonumber \\[0.3cm]
&& \left. \left. \textrm{artanh} \left[ \frac{r_B | \cos\theta |}{\sqrt{D^2 \sin^2\theta + r_B^2}} \right]\right) -\xi_1(\theta,D) 
\right], \\[0.5cm]
I_2(\theta,D) = &&\frac{\rho_{B}\ r_{B}}{2} \left[ \frac{r_{B}}{\sqrt{D^2 \sin^2\theta + r_B^2}} 
\left(  \frac{\pi}{2} + \textrm{artanh} \left[ \frac{r_B | \cos\theta |}{\sqrt{D^2 \sin^2\theta + r_B^2}}  \right] \right) \right. 
\nonumber \\[0.3cm]
&&-\xi_2(\theta,D) \Bigg].
\end{eqnarray}
Here,
\begin{eqnarray}
\xi_1(\theta,D)=\left\{
\begin{array}{ll}
\frac{r_B}{\sqrt{D^2 \sin^2\theta - r_B^2}}\ \arctan\left[ \frac{|\cos\theta | \sqrt{D^2 \sin^2\theta - r_B^2}}{D \sin^2\theta + r_B}\right] 
& \ \ \ D \sin \theta > r_B \\[0.3cm]
\sqrt{\frac{D-r_B}{D+r_B}}  & \ \ \ D \sin \theta = r_B \\[0.4cm]
\frac{r_B}{\sqrt{r_B^2-D^2 \sin^2\theta}}\ \textrm{artanh}\left[ \frac{|\cos\theta | \sqrt{r_B^2-D^2 \sin^2\theta}}{D \sin^2\theta + r_B} 
\right] & \ \ \ D \sin \theta < r_B 
\end{array}
\right.
\end{eqnarray}
and 
\begin{eqnarray}
\xi_2(\theta,D)=\left\{
\begin{array}{ll}
\frac{r_B}{\sqrt{D^2 \sin^2\theta - r_B^2}}\ \arctan\left[ \frac{\sqrt{D^2 \sin^2\theta - r_B^2}}{r_B}\right] 
& \ \ \ D \sin \theta > r_B \\[0.5cm]
1  & \ \ \ D \sin \theta = r_B \\[0.3cm]
\frac{r_B}{\sqrt{r_B^2-D^2 \sin^2\theta}}\ \textrm{artanh}\left[ \frac{\sqrt{r_B^2-D^2 \sin^2\theta}}{r_B} 
\right] & \ \ \ D \sin \theta < r_B 
\end{array}
\right..
\end{eqnarray}

\section{Required exposure for larger minimum energy}
\label{RExpo}

\begin{figure}[!h]
\centerline{\includegraphics[width=7.7cm]{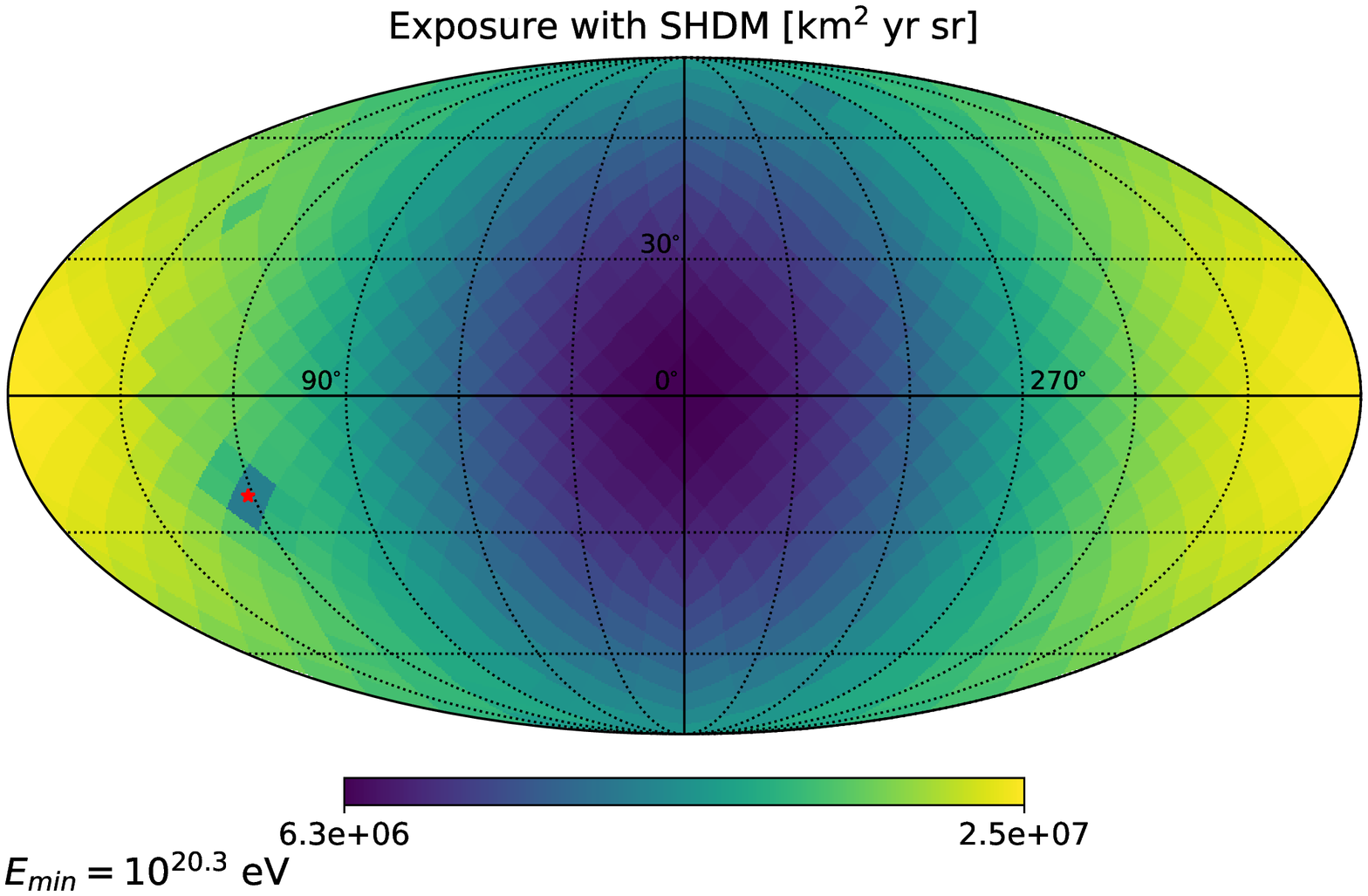}
\includegraphics[width=7.7cm]{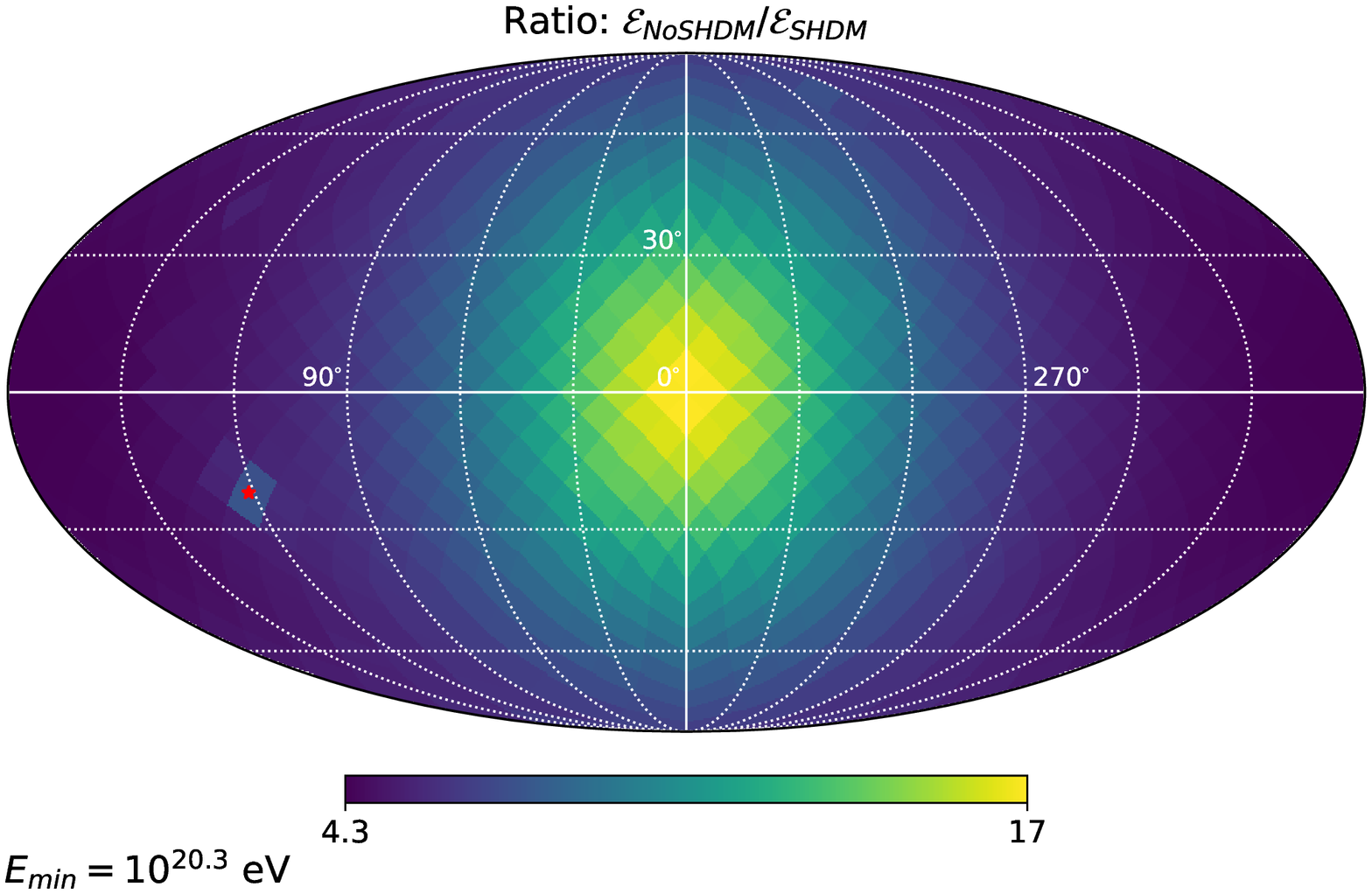}}
\caption{Left panel: Exposure required to observe at least one event in each pixel with 0.95 probability including all contributions
considered. Right panel: Ratio between the exposure required to observe at least one event in each pixel with 0.95 probability without 
and with the inclusion of the SHDM component. The minimum energy is $E_{min}=10^{20.3}$ eV and the red star corresponds to the position 
of Andromeda. \label{SkyMapExp20.3}}
\end{figure}

The left panel of Fig.~\ref{SkyMapExp20.3} shows the exposure required to observe at least one event in each pixel for 
$E_{min}=10^{20.3}$ eV and $p_0=0.95$ and for the case in which the contribution from SHDM is non negligible. The right 
panel of the figure shows the ratio between the exposure required to observed at least one event in each pixel with 0.95 
probability in the scenario without a SHDM contribution and in the one with a non negligible SHDM contribution. 
%

%
%
%
%

\acknowledgments
A.~D.~S.~is member of the Carrera del Investigador Cient\'ifico of CONICET, Argentina. This work is 
supported by ANPCyT PICT-2015-2752, Argentina. The authors thank the members of the Pierre Auger Collaboration 
for useful discussions and R. Clay for reviewing the manuscript.



\end{document}